\documentclass[epsfig,floats,prl,twocolumn,showpacs,
superscriptaddress,preprintnumbers,floatfix]{revtex4-1}
\usepackage{graphicx}
\usepackage[latin2]{inputenc}
\usepackage{amsmath}
\usepackage{amssymb}
\usepackage{bm}
\usepackage{color}
\newcommand{\Fl} {\mathcal{F}(\ell)}
\newcommand{\R}  {\mathcal{R}}

\begin{document}

\title{Anomalous Diffusion of Self-Propelled Particles in Directed Random Environments}
\author{M.\ Reza Shaebani}
\email{shaebani@lusi.uni-sb.de}
\affiliation{Department of Theoretical Physics, Saarland 
University, D-66041 Saarbr\"ucken, Germany}
\author{Zeinab Sadjadi}
\affiliation{Department of Theoretical Physics, Saarland 
University, D-66041 Saarbr\"ucken, Germany}
\author{Igor M.\ Sokolov}
\affiliation{Institut f\"ur Physik, Humboldt-Universit\"at 
zu Berlin, Newtonstrasse 15, 12489 Berlin, Germany}
\author{Heiko Rieger}
\affiliation{Department of Theoretical Physics, Saarland 
University, D-66041 Saarbr\"ucken, Germany}
\author{Ludger Santen}
\affiliation{Department of Theoretical Physics, Saarland 
University, D-66041 Saarbr\"ucken, Germany}
\date{\today}

\begin{abstract}
We theoretically study the transport properties of 
self-propelled particles on complex structures, such 
as motor proteins on filament networks. A general 
master equation formalism is developed to investigate 
the persistent motion of individual random walkers, 
which enables us to identify the contributions of key 
parameters: the motor processivity, and the anisotropy 
and heterogeneity of the underlying network. We prove 
the existence of different dynamical regimes of 
anomalous motion, and that the crossover times between 
these regimes as well as the asymptotic diffusion 
coefficient can be increased by several orders of 
magnitude within biologically relevant control parameter 
ranges. In terms of motion in continuous space, the 
interplay between stepping strategy and persistency 
of the walker is established as a source of anomalous 
diffusion at short and intermediate time scales.
\end{abstract}


\maketitle

Anomalous transport of self-propelled particles in 
biological environments has received much recent 
attention \cite{Reviews}. Of particular interest 
is the active motion of motor proteins along cytoskeletal 
filaments, which makes long-distance intracellular 
transport feasible \cite{Ross08}. The structural 
asymmetry of filaments results in a directed motion 
of motors with an effective \emph{processivity}, 
denoting the tendency to move along the same filament. 
The processivity depends on the type of motor and 
filament \cite{ProcesivityRefs1} and it is strongly 
influenced by the presence of specific proteins or 
binding domains \cite{Vershinin07,ProcesivityRefs2}. 
In the limit of small unbinding rates it has been 
shown \cite{Klumpp05} that a walker on simple lattice 
structures moves superdiffusively at short time 
scales followed by a normal diffusion at long times. 
Similar results were reported for single bead motion 
on radially-organized microtubule networks 
\cite{Caspi00}. However, for general polarized 
cytoskeletal networks, the influence of structural 
complexity and motor processivity on the transport 
properties is not yet well understood. In this Rapid 
Communication, we introduce a coarse-grained perspective 
to the problem and show that the interplay between 
anisotropy and heterogeneity of the network and 
processivity leads to a rich transport phase diagram 
at short and intermediate time scales. The crossover 
times between different regimes and the asymptotic 
diffusion constant can vary by orders of magnitude 
when tuning the key parameters. 

More precisely, a general analytical framework is 
developed to study persistent walks with arbitrary 
step-length and turning-angle distributions. We obtain 
an exact analytical expression for the dynamical 
evolution of the mean square displacement (MSD), 
displaying anomalous diffusion on varying time scales. 
The results can be also interpreted within the context 
of random motion in continuous space, e.g.\ in crowded 
biological media where the origin of subdiffusive motion 
is highly debated \cite{Subdiffusion3,Neusius09,
CTRWFBM,Jeon12b,Bronstein09,Thiel13}.

\begin{figure}[b]
\centering
\includegraphics[scale=0.41,angle=0]{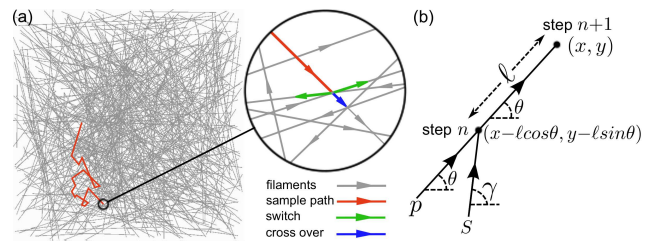}
\caption{(color online) (a) A typical sample trajectory 
(red lines) on a random filamentous structure. Inset: 
Possible choices at the nodes of the directed network. (b) 
Path of the walker during two consecutive steps, described 
by Eq.~(\ref{Eq:MasterEquation}).}
\label{Fig1}
\end{figure}

While subdiffusion in cytoplasm slows down the transfer 
of matter, it is beneficial for a variety of cellular 
functions \cite{Sereshki12-Guigas08,Golding06,Weigel11}, 
since they depend on the localization of the involved 
reactants. The appearance of subdiffusion has been 
traced back to the presence of particular elements 
e.g.\ viscoelasticity or traps in the environment, 
and the resulting motion is commonly characterized 
by comparing it to different mathematical models, e.g., 
\emph{fractional Brownian motion} \cite{FBM} which 
is a mean-zero Gaussian process with long-ranged 
(anti-)correlation of displacements, or \emph{continuous 
time random walk} \cite{CTRW} which assumes a finite 
variance of step lengths and a heavy tailed distribution 
of waiting times, as experienced e.g.\ by tracer 
particles in entangled actin filament networks \cite{Wong04} 
or among random energy traps \cite{Burov07}. More recent 
studies even suggest the coexistence of both scenarios 
\cite{CTRWFBM,Weigel11}. Here we verify that even in 
the absence of traps, viscoelasticity, overcrowding, 
etc., in the nature of the environment, the particle 
may still experience an anomalous motion within 
experimentally relevant time scales due to its 
stepping strategy. A specific strategy manifests itself 
for instance by the density, strength, and spatial 
obstacle arrangement and/or by external drives. We 
clarify the role of the variance of step lengths, the 
correlation between consecutive switching angles, and 
the persistency of the walker in displaying various 
types of motion. 
 
\begin{figure}[t]
\centering
\includegraphics[scale=0.4,angle=0]{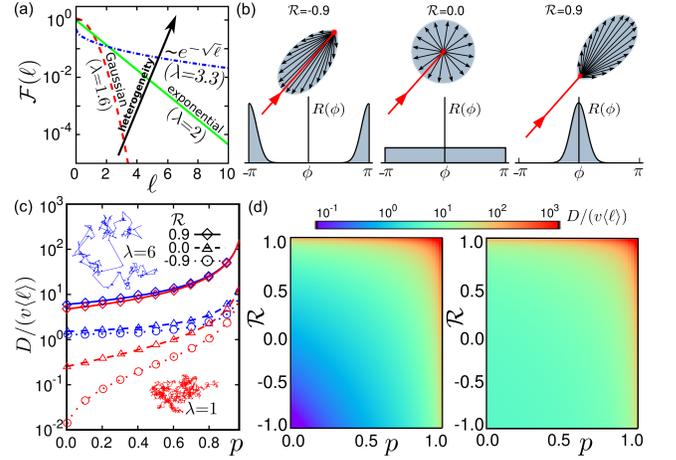}
\caption{(color online). (a) Evolution of $\lambda$ 
with $\Fl$. (b) Examples of $R(\phi)$ and their 
corresponding anisotropy parameter $\R$. Insets: 
Black arrows show the possible directions of motion 
at the next step, with length being proportional 
to the probability. (c) Asymptotic diffusion 
coefficient vs $p$ at different $\R$, from simulations 
(symbols) and analytical predictions via Eq.~(\ref{Eq:D}) 
(curves). The red (blue) color corresponds to 
$\lambda=1\,(6)$. Insets: Typical trajectories at 
$p{=}\R{=}0$. (d) 2D profiles of $D$ in $p{-}\R$ 
plane for $\lambda{=}1$ (left) and $\lambda{=}6$ 
(right).}
\label{Fig2}
\end{figure}

{\it Model and analytical solution.---}
We focus on the network interpretation in the following, and 
consider the motion of a walker on a randomly cross-linked 
network of polarized filaments [Fig.~\ref{Fig1}(a)]. The 
structural properties are characterized by probability 
distributions $R(\phi)$ for the angles $\phi$ between 
intersecting filaments, and $\Fl$ for the segment lengths 
$\ell$ between neighboring intersections \cite{remark1}. 
Here we study a 2D network (extension to 3D is 
straightforward \cite{Shaebani13}), and describe 
the motion of the particle by a Markovian process in 
discrete time and denote the probability density to be 
at position $(x,y)$ with a direction $\theta$ at time 
step $n$ by $P_n(x,y|\theta)$, whose dynamical evolution 
is defined by the master equation
\begin{eqnarray}
\begin{aligned}
&P\!\!_{_{n+1}}\!(x,y|\theta)=p\!\!\int\!\!\!d\ell\,
\Fl\,P\!\!_{_n}\!\big(x{-}\ell\text{cos}(\theta),y{-}
\ell\text{sin}(\theta)\big|\theta\big)\\
&{+}s\!\!\int\!\!\!d\ell\,\Fl\!\!\int_{-\pi}^{\pi}
\!\!\!\!\!d\gamma\,R(\theta{-}\gamma)\,P\!\!_{_n}\!
\big(x{-}\ell\text{cos}(\theta),y{-}\ell\text{sin}
(\theta)\big|\gamma\big).
\end{aligned}
\label{Eq:MasterEquation}
\end{eqnarray}
The probability of motion without changing the 
direction, $p$, represents the processivity of the motor, 
while $s{=}1{-}p$ describes a directional change [see 
Fig.~\ref{Fig1}(b)]. While it is quite difficult to 
obtain an explicit analytical expression for 
$P_{n}(x,y|\theta)$ from Eq.~(\ref{Eq:MasterEquation}), 
it is possible to evaluate the moments of the 
displacement using an analytical Fourier{\it--}Z-transform 
technique \cite{Shaebani13,Sadjadi08}. We obtain 
the following exact expression for the MSD \cite{remark2}
\begin{equation}
\begin{aligned}
\frac{\langle r^2\rangle\!_{_n}}{\langle\ell\rangle^2}
=\lambda\,n{+}\!\!\!\!\!\sum_{m{=}\pm1}\!\!\!\frac{p{+}
\R\!_{_m}{-}p\R\!_{_m}}{(1{-}p)(1{-}\R\!_{_m})}\bigg[n
{+}\frac{(p{+}\R\!_{_m}{-}p\R\!_{_m})^n{-}1}{(1{-}p)
(1{-}\R\!_{_m})}\bigg],
\end{aligned}
\label{Eq:X2}
\end{equation}
where $\R\!_{_{m}}$ is the Fourier transform of 
the intersection angle distribution, $\R\!_{_{m}}
{=}\int_{-\pi}^{\pi}\text{d}\phi\;e^{\text{im}\phi}
R(\phi)$, and $\lambda{=}\langle\ell^2\rangle/\langle
\ell\rangle^2$ is the relative variance of $\Fl$ which 
quantifies the heterogeneity of the network 
[Fig.~\ref{Fig2}(a)]. $\R\!_{_{m}}{\in}[-1,1]$ 
quantifies the correlation between the arrival 
direction before a switch and the final direction 
after it, thus, representing the anisotropy of the 
network. $\R\!_{_{m}}{=}0$ corresponds to the 
uniform case, and negative (positive) values of 
$\R\!_{_{m}}$ to an increased probability for 
motion in the near backward (forward) directions 
[Fig.~\ref{Fig2}(b)]. Although our method allows us 
to handle an arbitrary function $R(\phi)$, here we 
consider only symmetric distributions with respect 
to the arrival direction ($\R\!_{_{-1}}{=}\R\!_{_{+1}}
{\equiv}\R$), as it is usually the case in biological 
applications. The results remain valid in three 
dimensions for a cylindrical symmetry of $R(\phi)$. 
In the simple case of walking with left-right 
symmetry and without persistency, Eq.~(\ref{Eq:X2}) 
reduces e.g.\ to a model for cell migration along 
surfaces \cite{Nossal74}. 

{\it Asymptotic behavior.---} 
In the limit of $n {\rightarrow}\infty$ the terms 
with $(p{+}\R{-}p\R)^n$ in Eq.~(\ref{Eq:X2}) vanish 
since $|p{+}\R{-}p\R|{\leq}1$. Hence the linear term 
dominates and the motility becomes purely diffusive. 
The asymptotic diffusion constant is given by 
\begin{equation}
\begin{aligned}
D=\frac{1}{4}v\langle\ell\rangle \bigg[\lambda+
\frac{2(p{+}\R{-}p\R)}{(1{-}p)
(1{-}\R)}\bigg],
\end{aligned}
\label{Eq:D}
\end{equation}
with $v$ being the average motor velocity. Figures 
\ref{Fig2}(c),(d) show that $D$ varies by several orders 
of magnitude by varying the anisotropy and heterogeneity 
measures, $\R$ and $\lambda$, and the processivity $p$. 
The results of extensive Monte Carlo simulations for 
persistent walk on structures obtained from the same 
$R(\phi)$ and $\Fl$ distributions agree perfectly with 
analytical predictions. The linearly additive contribution 
of $\lambda$ in Eq.~(\ref{Eq:D}) is negligible when 
$p,\R{\rightarrow}1$ but dominates for $p,\R{\ll}1$ 
and large $\lambda$. Intuitively, a broader 
distribution $\Fl$ corresponds to a larger 
asymptotic diffusion coefficient $D$ (e.g.\ 
$D_\text{exp}/D_\text{Gaussian}{\simeq}1.27$ 
for $p{=}\R{=}0$). For a walker on square 
lattice with $p{=}0$, Eq.~(\ref{Eq:D}) reduces to 
$D{=}\frac14 v\langle\ell\rangle$ \cite{Klumpp05}. 
The negative values of $\R$ represent anticorrelation 
between consecutive switching angles. A pure localization 
($D{=}0$) happens when $\R{=}-1$, $p{=}0$, and $\lambda{=}1$. 

Motor proteins are highly flexible to turn even up to 
$150^\circ$ at intersections \cite{ProcesivityRefs1}, 
and have a typical velocity $v{\sim}1\mu\text{m/s}$ 
\cite{Ali08}. $\R$ may vary from $0$ for a random 
actin-filament structure to $1$ for radially-organized 
microtubule networks, and processivity can be tuned at 
least within the range $0.1{-}0.7$ \cite{ProcesivityRefs2,
Vershinin07}. Hence, for an actin network with an average mesh 
size $\langle\ell\rangle{\sim}100\,\text{nm}$ \cite{Wong04},
$D$ may vary by $3$ orders of magnitude from $10^{-2}$ 
to $10\mu\text{m}^2\text{/s}$. 

\begin{figure}[t]
\centering
\includegraphics[scale=0.6,angle=0]{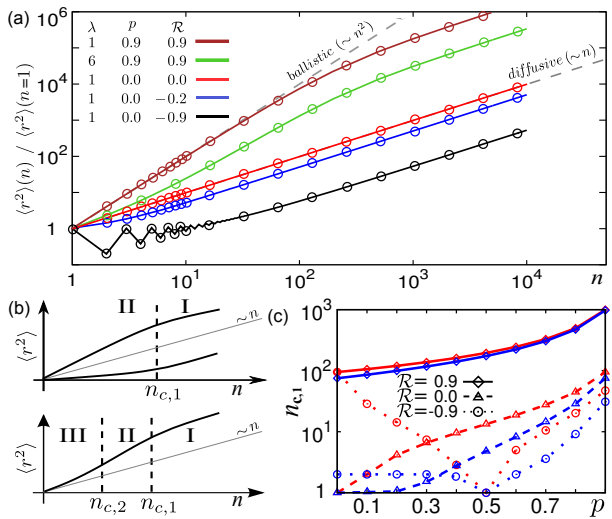}
\caption{(color online). (a) MSD vs $n$ at different 
$\lambda$, $p$, and $\R$, from simulations (symbols) 
and theory via Eq.~(\ref{Eq:X2}) (curves). (b) Schematic 
representation of the possible regimes of motion and 
the crossover times. (c) The crossover time $n_\text{c,1}$ 
vs $p$ for different values of $\R$, for $\lambda{=}1$ 
(red) and $\lambda{=}6$ (blue).}
\label{Fig3}
\end{figure}

{\it Different regimes of motion.---} 
A random walker with a constant step length and uniform 
rotation angle ($p{=}\R{=}0$) conceivably displays 
normal diffusion at all time scales. In the general 
case, however, we predict a rich variety of the MSD 
profiles, as shown in Fig.~\ref{Fig3}. The profile 
shapes strongly depend on the choice of $\lambda$, 
$p$, and $\R$. An oscillatory dynamics emerges at 
large negative values of $\R$, where the motion is 
strongly antipersistent and the particle hops 
frequently back and forth without a significant 
net motion. Such behavior is observed for the 
antipersistent motion of paramagnetic colloids in 
a periodically switching magnetic potential 
\cite{Tierno12}.

Although the asymptotic dynamics of a system described by 
Eq.~(\ref{Eq:X2}) is diffusive, the corresponding diffusion 
scaling $\langle\textbf{r}^2\rangle{\propto}Dn$ might 
be observable only on time- and length scales which are 
experimentally not accessible. We denote the asymptotic 
regime by I and the preceding regime (oscillatory, sub- 
or super-diffusive) by II [Fig.~\ref{Fig3}(b)]. The smooth 
crossover from regime II to I occurs at a characteristic 
time scale $n_{\text{c},1}$ which strongly varies with the  
parameter values [see Fig.~\ref{Fig3}(a)]. $n_{\text{c},1}$ 
is identified by balancing the term linear in $n$ in 
Eq.~(\ref{Eq:X2}) and the non-linear contribution. 
Figure~\ref{Fig3}(c) shows that $n_{\text{c},1}$ varies over 
several orders of magnitude. For an actin network, the 
time scale for motors to travel between filament junctions 
is ${\sim}0.1\,\text{sec}$ (see above), and thus $n_{\text{c},1}$ 
ranges from less than $10^{-1}$ up to more than $10^{2}$ 
seconds for the motion of motor proteins on actin filament 
networks. Experimentally, it is often difficult to perform 
measurements in a time window which is wide enough to ensure
 that all different regimes of motion are realized. 

$n_{\text{c},1}$ decreases with increasing $\lambda$, 
i.e.\ heterogeneity extends the diffusive regime I to 
smaller times. While $n_{\text{c},1}$ increases 
monotonically with $p$ for $\R{\ge}0$, it exhibits a 
nonmonotonic dependence for $\R{<}0$, with the minimum 
value $n_{\text{c},1}{=}1$ at a processivity 
$p\!_{_\text{min}}{\simeq}\R/(\R{-}1)$. When $\R{>}0$, 
$p$ and $\R$ cooperate in enhancing net forward motion which 
results in a superdiffusive dynamics in regime II. Therefore, 
increasing $p$ while $\R{\ge}0$ leads to stronger 
superdiffusion and postpones the transition to regime 
I, reflected by a monotonic increase of $n_{\text{c},1}$. 
In contrast, $p$ and $\R$ compete when $\R{<}0$, which 
may lead to a variety of anomalous diffusion scenarios, 
depending on their relative importance. With increasing 
$p$ from zero at a negative $\R$, the anticorrelation 
of rotation angles initially dominates the dynamics and 
causes oscillations or subdiffusion. However, the role 
of processivity $p$ becomes gradually more pronounced 
and a transition from sub to superdiffusion happens in 
regime II at the turning point $p\!_{_\text{min}}$. 
Thus the varying strength and type of anomaly are 
the reasons behind the nonmonotonic behavior of 
$n_{\text{c},1}$. 

\begin{figure}[t]
\centering
\includegraphics[scale=0.38,angle=0]{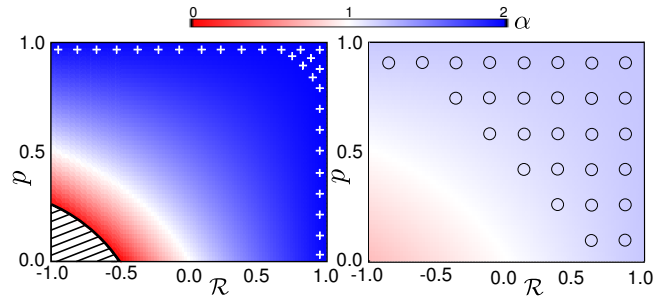}
\caption{(color online). 2D profiles of the phase 
diagram of short time dynamics in the $p{-}\R$ plane 
for $\lambda{=}1$ (left) and $\lambda{=}6$ (right). 
The color intensity reflects the magnitude of 
anomalous exponent $\alpha$, with red (blue) meaning 
sub- (super-) diffusion. The white plus symbols 
and hatched lines denote the ballistic and oscillatory 
subdomains. The circles mark the regions of the $p{-}\R$ 
parameter space where the slope of $\langle\textbf{r}^2
\rangle$ initially grows with time (regime III). The 
blue-red (red-oscillatory) interface is specified 
by $p{=}\frac{\R}{\R{-}1}$ ($p{+}\R{-}p\R{=}-\lambda{/}2$).}
\label{Fig4}
\end{figure}

{\it Short-time dynamics.---} We determine an 
effective anomalous exponent $\alpha{=}1{+}\ln(1{+}
\frac{p{+}\R{-}p\R}{\lambda})/\ln2$ by fitting the 
initial $t$-dependence of the MSD 
to a power-law $\langle\textbf{r}^2\rangle
{\sim}t^\alpha$.  Fig.~\ref{Fig4} summarizes the results 
of the anomalous short-term motion in a phase diagram 
in the ($p$,$\R$) space. The larger values of $\lambda$ 
imply effective exponents that are closer to $1$ as 
a consequence of the increasing role of the linear 
terms in $\langle\textbf{r}^2\rangle$. Above the 
threshold heterogeneity $\lambda_c{=}2$, the oscillatory 
phase no longer exists.

Finally, a closer look at $\langle\textbf{r}^2\rangle$ 
profiles reveals that in some of the superdiffusive 
cases the initial growth of $\langle\textbf{r}^2\rangle$ 
gets accelerated [regime III in Fig.~\ref{Fig3}(b)], 
although the curvature changes sign later, i.e. a 
crossover to regime II eventually happens. To explain 
this, we expand $\langle\textbf{r}^2\rangle$ around 
$n{=}1$ for the parameter values corresponding to 
the blue regions in Fig.~\ref{Fig4}, and find that 
the main contributions originate from the first and 
second order terms in $n$. From the competition between 
these terms we determine a characteristic (short) 
time scale 
$n_{\text{c},2}(\lambda,\R,p){\simeq}2\big[
(\lambda{+}\mathcal{A})(1{-}p)(1{-}\R){+}
\mathcal{A} \;\text{ln}\mathcal{B}\big]/\big[
\mathcal{A}\;(\text{ln}\mathcal{B})^2\big],$
where $\mathcal{B}{=}p{+}\R{-}p\R$ 
and $\mathcal{A}{=}2\mathcal{B}/[(1{-}p)(1{-}\R)]$. 
At times $n{\ll}n_{\text{c},2}$, the linear term (${\sim}n$) 
plays the dominant role, thus, keeping the initial slope 
slightly above $1$ (regime III), while at 
$n_{\text{c},2}{\ll}n$ the second order term (${\sim}n^2$) 
dominates and the slope increases. By varying the control 
parameters, $n_{\text{c},2}$ can be pushed towards or 
away from $n{=}1$. In the limit of $n_{\text{c},2}
{\rightarrow}1$, regime III disappears and the slope of 
$\langle\textbf{r}^2\rangle$ initially starts with the 
extremum value (regime II) and ends up with $1$. In the 
case that $1{\ll}n_{\text{c},2}$, all regimes exist. It 
turns out that the crossover time remains around 
$n_{\text{c},2}{\simeq}1$ for $\lambda{=}1$, i.e.\ 
regime III does not exist. However, with increasing 
$\lambda$, this regime appears for large values of 
$p$ and $\R$, and gradually spans the whole 
superdiffusive (blue) subdomain for large values of 
$\lambda$.   

\begin{figure}[t]
\centering
\includegraphics[scale=0.4,angle=0]{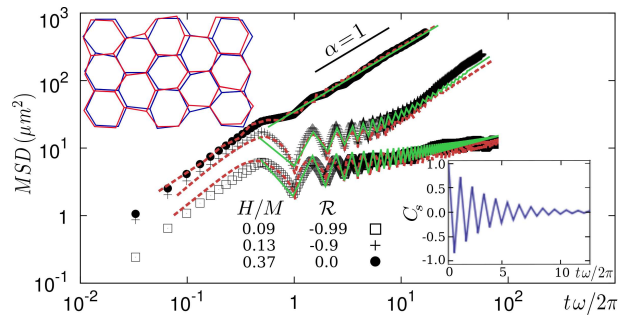}
\caption{(color online). MSD versus time for a 
paramagnetic tracer particle in the two-state flashing 
potential. The analytical (solid lines) and simulation 
(dashed lines) fits are compared with experimental data 
(symbols) taken from Fig.~2 of \cite{Tierno12}. 
Insets: (top) Schematic picture of two consecutive 
disordered lattices in simulations with $\delta{=}10\%$. 
(bottom) The normalized step-step correlation function 
$C_s$ vs.\ time.}
\label{Fig5}
\end{figure}

{\it Comparison with experimental data.---} Besides 
living organisms, the twofold interpretation of our 
methodology extends the applicability range of the 
results to the motion of self-propelled particles in 
continuous space in other realizations such as driven 
granular systems \cite{GranularRefs} where the 
stepping properties of particles can be tuned by 
means of an external drive source and internal 
conditions \cite{remark1}. To validate the theoretical 
predictions, here we compare the analytical results 
with experiments in a nonliving system \cite{Tierno12}, 
where the accelerated motion of a paramagnetic tracer 
particle on a ferrite garnet film is controlled 
externally. The film displays a 2D array of magnetic 
bubbles with out-of-plane magnetization $M$, sitting 
on a hexagonal lattice of constant $L{=}11.6\mu\text{m}$. 
An external magnetic field switches between $\pm{H}$ 
with frequency $\omega$, which induces a certain degree 
of disorder in the shape and arrangement of the magnetic bubbles 
\cite{Soba08}. The resulting dynamic disorder (i.e.\ 
not reproducible after each cycle) influences the 
antipersistent motion of the tracer particle. In the 
absence of additional sources of anomalous behavior, 
such a labyrinthine environment provides an opportunity 
to examine the isolated impact of stepping strategy, 
which is tuned via remote control. For comparison, we 
also carried out simulations of motion on a dynamic 
disordered hexagonal lattice of synchronous flashing 
magnetic poles, which would closely mimic the experiment 
\cite{Shaebani13}. At each time step $\Delta t{=}\pi{/}
\omega$, the new position of each node is randomly 
chosen within a circle of radius $L{\cdot}\delta$ 
around the corresponding node of the ordered hexagonal 
lattice ($\delta$ reflects the amount of disorder). 
The resulting step-length distribution $\Fl$ 
of the tracer particle is fitted by a Gaussian centered 
at the size of the Wigner-Seitz cell, $L{/}2$, with 
a standard deviation e.g.\ $\sigma{\simeq}0.05L$ for 
$\delta{=}10\%$. With increasing $\delta$, the 
rotation-angle distribution $R(\phi)$ evolves from a 
single-peaked function at $\phi{=}\pi$ towards multiple 
peaks at $\phi{=}\pi,{\pm}2\pi{/}3,{\pm}\pi{/}3$. By 
tuning $\R$ in theory or the lattice disorder 
$\delta$ in simulations, the comparison with experimental 
data shows a remarkable agreement, as shown in 
Fig.~\ref{Fig5}. Notably, the frequency of oscillations 
is correctly reproduced and, moreover, one can derive 
the step-step correlations as $C_s(t){=}\R^{t
\omega{/}\pi}$, which gives rise to the same behavior 
reported in \cite{Tierno12}.

{\it Discussion.---} To establish the interplay between 
processivity and structural properties of cytoskeleton 
as a source of anomalous motion, we have taken only angular 
correlations into account in the present model, which 
defines a characteristic correlation scale beyond which 
the motion is diffusive (see the inset of Fig.~\ref{Fig5}). 
However, the formalism can be generalized in different 
ways, e.g.\ by introducing a long-range (anti-)cross-correlation 
between the angular and step-length distributions, leading 
to a stationary increment as observed in viscous environments. 
Moreover, intermittent walks can be considered, using a 
set of coupled master equations. By handling different 
modes of motility (e.g.\ binding and unbinding of motor 
proteins, or a combined waiting-running motion as 
experienced in the presence of traps or in overcrowded 
environments), one can obtain exact expressions for the 
time evolution of arbitrary moments of displacement. 
The possibility of handling arbitrary stepping roles in 
our approach, which is not feasibly available in other 
approaches, provides a unique opportunity to study more 
complex situations and develop models with predictive power. 

We verified that self-propelled particles display a wide 
range of different types of motion on complex structures. 
We disentangled the combined effects of processivity 
and structural properties of the underlying network on 
transport properties. The method relates the microscopic 
details of the transport network or the characteristics 
of the particle dynamics to macroscopically observable 
transport coefficients such as the diffusion coefficient, 
without using phenomenological or purely mathematical 
models. The results point to new strategies to unravel 
the structural complexity of filamentous networks or 
non-biological realizations such as porous media by 
monitoring the motion of tracer particles which 
perform a random walk on such environments. 

We thank P. Tierno for supplying the experimental data. 
This work was supported by DFG within SFB 1027 (projects 
A7, A3), and BMBF (FKZ 03X0100C).


\begin{thebibliography}{10}
\bibitem{remark1} $R(\phi)$ and $\Fl$ can be interpreted 
as rotation-angle and step-length probabilities in the 
continuous space.
\bibitem{remark2} This particular result can also 
be obtained by adapting other analytical techniques 
(see e.g. \cite{Nossal74,Kareiva83}), yet, our method is directly 
applicable to a large class of persistent random 
walk problems, which are out of scope for the other methods.
\end{thebibliography}
\end{document}